# Very forward calorimeters for future electron-positron colliders

Ivan Smiljanic[a,1]

[a]Vinca Institute of Nuclear Sciences, National Institute of the Republic of Serbia, University of Belgrade, Belgrade, Serbia

**Abstract:** Detectors at future e+e- colliders need special calorimeters in the very forward region for a fast estimate and precise measurement of the luminosity, to improve the hermeticity and mask the central tracking detectors from backscattered particles. In our concept, two compact calorimeters are foreseen, LumiCal and BeamCal. Both are designed as sandwich calorimeters with very thin sensor planes to keep the Molière radius small, facilitating such the measurement of electron showers in the presence of background. Silicon sensor prototypes and dedicated FE ASICs have been developed and produced. The ASICs match the timing and dynamic range requirements. In the recent beam tests, a multi-plane compact prototype was equipped with thin sensor planes fully assembled with the new readout electronics and installed in 1 mm gaps between tungsten plates of one radiation length thickness. The latest status of the calorimeter prototype development will be presented, including selected performance results, obtained in a 5 GeV electron beam at DESY, as well as the expected performance obtained from simulation.

## 1. Introduction

In the forward region of future high energy linear $e^+e^-$ colliders [1, 2], two specialized calorimeters are foreseen: luminosity calorimeter - LumiCal for the integrated luminosity measurement and BeamCal for fast luminosity estimate and the beam parameters control. LumiCal and BeamCal are cylindrical electromagnetic calorimeters designed as sensor-tungsten sandwich; the absorber thickness is 3.5 mm, corresponding to one-radiation length. The longitudinal structure of both calorimeters consists of layers of absorber interspersed with very thin detector planes. Several solid-state sensor materials [3], including GaAs [4], diamond [5] and single-crystal sapphire [6], along with conventional silicon diode sensors, are being assessed for radiation tolerance. The main parameters of the LumiCal and BeamCal in ILD and CLICdet detector concepts are shown in Table 1.

In this paper, the latest developments towards the compact LumiCal prototype calorimeter as well as performance highlights from test-beam campaigns will be presented.

**Table 1.** Main parameters of the LumiCal and BeamCal in ILD and CLICdet detector concepts

|  | Parameters | ILD | CLICdet |
|---|---|---|---|
| **LumiCal** | geometrical acceptance [mrad] | 31-77 | 38-110 |
|  | Fiducial acceptance [mrad] | 41-67 | 44-80 |
|  | z from IP [mm] | 2480 | 2539 |
|  | number of layers (W+Si) | 30 | 40 |
| **BeamCal** | geometrical acceptance [mrad] | 5-40 | 10-40 |
|  | z from IP [mm] | 3200 | 3181 |
|  | number of layers (W+Si) | 30 | 40 |

## 2. LumiCal prototype

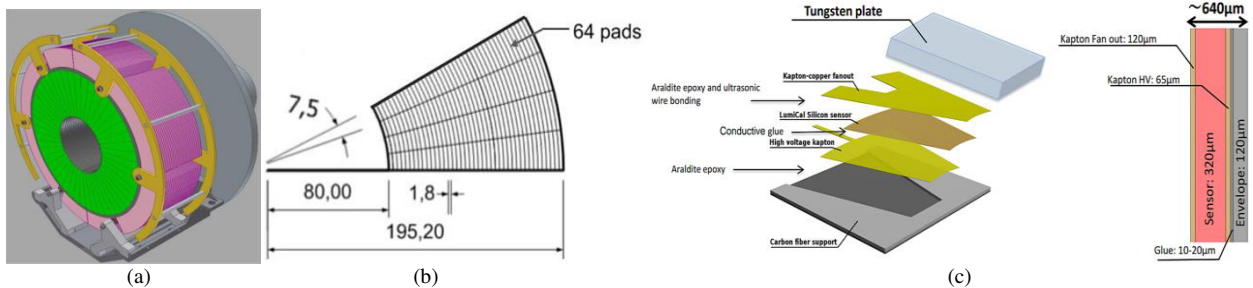

**Figure 1.** (a) Model of LumiCal; (b) Sector of the LumiCal Si sensor; (c) Detector plane assembly

The combination of Si and W for the construction of the detector allows the assembly of a very compact calorimeter made up of compact active layers with small cell size (high granularity) in the transverse and longitudinal directions (Figure 1a). The sensor is made of a 320 μm thick high resistivity n-type silicon wafer. It has the shape of a sector of a 30° angle, with inner and outer radii of the sensitive area of 80 mm and 195.2 mm, respectively (Figure 1b). It comprises four sectors with 64 p-type pads of 1.8 mm pitch. The silicon sensor was glued with epoxy to a fan-out made of flexible

---

[1] On behalf of FCAL Collaboration

Kapton-copper foil. Ultrasonic wire bonding was used to connect conductive traces on the fan-out to the sensor pads. A 70 μm thick Kapton-copper foil glued on the back side of the sensor with a conductive glue supplied the high voltage. For good mechanical stability, the assembled sensor was embedded in a carbon fiber envelope. The fully assembled detector plane developed for this study (Figure 1c) has a total thickness of about 650 μm. LumiCal is a Si-W electromagnetic sandwich calorimeter designed to measure the integrated luminosity with a precision better then $10^{-3}$ for ILC and $10^{-2}$ for CLIC. We have built a prototype of LumiCal as a very compact calorimeter with ultra-thin sensors and tested it.

## 3. Test beam campaigns

The FCAL Collaboration performed several tests of the LumiCal prototype between 2014 and 2020. We started with 4 fully equiped planes in the proton beam at CERN, and for the last campaign we tested 15 planes (3 of them equipped with the new FLAME readout) in $e^-$ beam with energies between 1 GeV and 5 GeV at DESY-II Synchrotrone. Main setups used in tests are shown on Figure 2.

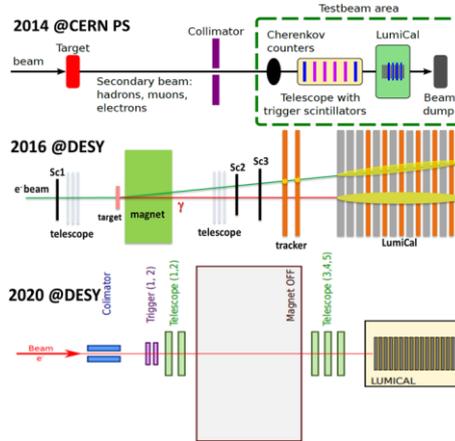

**Figure 2.** Setups used in test beam campaigns 2014-2020.

Main goals of those tests were measurements of the transversal profile and the longitudinal propagation of the shower in different setups and, consequently, measurement of Moliere radius.

For the last test-beam campaign, new readout electronics, FLAME (FcaL Asic for Multiplane rEadout) [3], has been developed and successfully tested. The ASIC for FLAME was designed in 130 nm CMOS technology and matches the timing requirements of a future linear collider, and has a larger dynamic range compared with the previously used APV-25 chip board.

## 3. Results

Figure 3. shows preliminary results of the calorimeter response to 5 GeV electron beam readout with the FLAME readout. Using the same approach as in [Halina], the value of effective Moliere radius is estimated to 10.1 mm. On Figure 4, the longitudinal shower profile is shown. Figure 4 (a) is obtained from data, by fitting the convolution of Landau and Gauss functions to the energy spectrum for each pad. Blue dots and line show the mean values and red the most probable value (MPV). Figure 4 (b) shows the result of the Monte Carlo simulation. There's a good agreement between the Monte Carlo and MPV. Both pictures show that the maximum of the longitudinal shower is reached at the $7^{th}$ plane ($7X_0$), which is in agreement with results from previous test-beam campaigns.

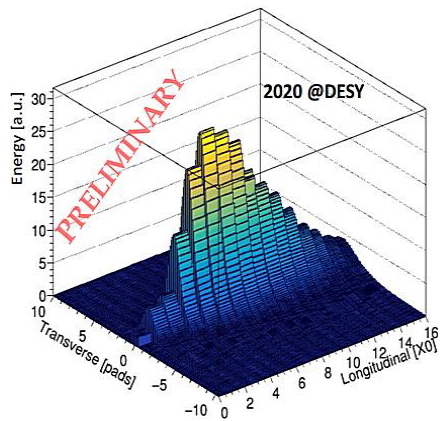

Figure 3. A lego plot of the transverse profile for each layer from the beam-test data.

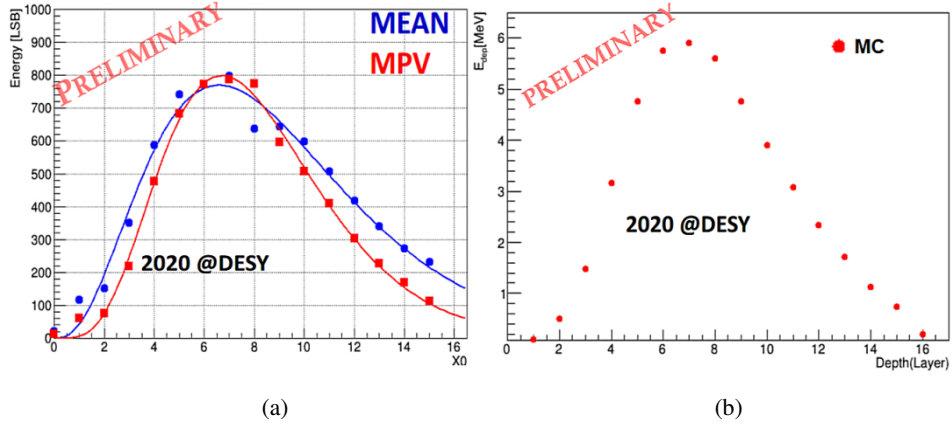

(a)                  (b)

**Figure 4.** A longitudinal shower profile in the LumiCal. (a) – test beam data; (b) – Monte Carlo.

## 4. Conclusion

With the presented results, we have proven that design of the compact calorimeter with ultra-thin sensors is possible. Analysis of data and Monte Carlo from the last test-beam campaign is still ongoing and the new campaigns are planned. The FCAL Collaboration continues to work on forward calorimetry for future Higgs factories, as well as for other experiments, such as LUXE.

## Acknowledgments

This activity was partially supported by the Romanian UEFISCDI agency under grant no. 16N/2019. These studies were partly supported by the Israel Science Foundation (ISF), Israel German Foundation (GIF), the I-CORE program of the Israel Planning and Budgeting Committee, Israel Academy of Sciences and Humanities, by the National Commission for Scientific and Technological Research (CONICYT - Chile) under grant FONDECYT1170345, by the Polish Ministry of Science and Higher Education under contract nrs 3585/H2020/2016/2 and 3501/H2020/2016/2, by the Ministry of Education, Science and Technological Development of the Republic of Serbia and by the Science Fund of the Republic of Serbia through the Grant No. 7699827, HIGHTONE-P, by the United States Department of Energy, grant de-sc0010107, and by the European Union Horizon 2020 Research and Innovation programme under Grant Agreement no.654168 (AIDA-2020). The measurements leading to these results have been performed at the Test Beam Facility at DESY Hamburg (Germany), a member of the Helmholtz Association (HGF).